\documentclass[conference]{IEEEtran}
\IEEEoverridecommandlockouts
\usepackage{cite}
\usepackage{amsmath,amssymb,amsfonts}
\usepackage{algorithmic}
\usepackage{graphicx}
\usepackage{textcomp}
\usepackage{xcolor}
\usepackage{booktabs} % For formal tables
\usepackage{amssymb}
\usepackage{soul}
\usepackage{hyperref}
\def\BibTeX{{\rm B\kern-.05em{\sc i\kern-.025em b}\kern-.08em
    T\kern-.1667em\lower.7ex\hbox{E}\kern-.125emX}}

\usepackage{tikz}
\usetikzlibrary{calc}

\makeatletter
\newif\if@anonymize

\@anonymizetrue    % Uncomment to hide text
\if@anonymize
  \newcommand{\highlight@DoHighlight}{
    \fill [outer sep = -15pt, inner sep = 0pt, color=black]
          ($(begin highlight)+(0,8pt)$) rectangle ($(end highlight)+(0,-3pt)$) ;
  }

  \newcommand{\highlight@BeginHighlight}{
    \coordinate (begin highlight) at (0,0) ;
  }

  \newcommand{\highlight@EndHighlight}{
    \coordinate (end highlight) at (0,0) ;
  }

  \newdimen\highlight@previous
  \newdimen\highlight@current
  \newlength{\item@width}

  \DeclareRobustCommand*\anonymize{%
    \SOUL@setup
    \def\SOUL@preamble{%
      \begin{tikzpicture}[overlay, remember picture]
        \highlight@BeginHighlight
        \highlight@EndHighlight
      \end{tikzpicture}%
    }%
    \def\SOUL@postamble{%
      \begin{tikzpicture}[overlay, remember picture]
        \highlight@EndHighlight
        \highlight@DoHighlight
      \end{tikzpicture}%
    }%
    \def\SOUL@everyhyphen{%
      \discretionary{%
        \SOUL@setkern\SOUL@hyphkern
        \SOUL@sethyphenchar
        \tikz[overlay, remember picture] \highlight@EndHighlight ;%
      }{%
      }{%
        \SOUL@setkern\SOUL@charkern
      }%
    }%
    \def\SOUL@everyexhyphen##1{%
      \SOUL@setkern\SOUL@hyphkern
      \settowidth{\item@width}{##1}%
      \makebox[\item@width]{}%
      \discretionary{%
        \tikz[overlay, remember picture] \highlight@EndHighlight ;%
      }{%
      }{%
        \SOUL@setkern\SOUL@charkern
      }%
    }%
    \def\SOUL@everysyllable{%
      \begin{tikzpicture}[overlay, remember picture]
        \path let \p0 = (begin highlight), \p1 = (0,0) in \pgfextra
          \global\highlight@previous=\y0
          \global\highlight@current =\y1
        \endpgfextra (0,0) ;
        \ifdim\highlight@current < \highlight@previous
          \highlight@DoHighlight
          \highlight@BeginHighlight
        \fi
      \end{tikzpicture}%
      \settowidth{\item@width}{\the\SOUL@syllable}%
      \makebox[\item@width]{}%
      \tikz[overlay, remember picture] \highlight@EndHighlight ;%
    }%
    \SOUL@
  }
\else
  \newcommand{\anonymize}[1]{#1}
\fi
\makeatother

\begin{document}

% TODO: TITOLO PROVVISORIO DA CAMBIARE 
% Io nel titolo ci ficcherei da qualche parte multilayer...a quanto diceva manlio é la parte che puó essere considerata piú innovativa.
\title{Weak nodes detection in urban transport systems: Planning for resilience in Singapore
\thanks{ Luca Pappalardo has been partially funded by the European project SoBigData RI (Grant Agreement 654024). Gianni Barlacchi has been partially funded by the EIT project Cedus.}
}

% \author{\IEEEauthorblockN{1\textsuperscript{st} Michele Ferretti}
% \IEEEauthorblockA{\textit{Department of Geography} \\
% \textit{King's College London}\\
% London, United Kingdom \\
% michele.ferretti@kcl.ac.uk}
% \and
% \IEEEauthorblockN{2\textsuperscript{nd} Gianni Barlacchi}
% \IEEEauthorblockA{\textit{dept. name of organization (of Aff.)} \\
% \textit{name of organization (of Aff.)}\\
% City, Country \\
% email address}
% \and
% \IEEEauthorblockN{3\textsuperscript{rd} Luca Pappalardo}
% \IEEEauthorblockA{\textit{dept. name of organization (of Aff.)} \\
% \textit{name of organization (of Aff.)}\\
% City, Country \\
% email address}

% \and
% \IEEEauthorblockN{4\textsuperscript{th} Lorenzo Lucchini}
% \IEEEauthorblockA{\textit{ICT} \\
% \textit{Fondazione Bruno Kessler}\\
% Trento, Italy \\
% llucchini@fbk.eu}
% \and
% \IEEEauthorblockN{5\textsuperscript{th} Bruno Lepri}
% \IEEEauthorblockA{\textit{dept. name of organization (of Aff.)} \\
% \textit{name of organization (of Aff.)}\\
% City, Country \\
% email address}
% }

% ------- FORMATO AUTORI LUNGO (piu' di tre) -------------
% ---------------------------------------------------------
%         NOTA:   TOLTI PERCHE' DOUBLE-BLINDED
% ---------------------------------------------------------
% ---------------------------------------------------------

 \author{
 \IEEEauthorblockN{
     Michele Ferretti\IEEEauthorrefmark{1},
     Gianni Barlacchi\IEEEauthorrefmark{2}\IEEEauthorrefmark{3},
     Luca Pappalardo\IEEEauthorrefmark{4},
     Lorenzo Lucchini\IEEEauthorrefmark{2}\IEEEauthorrefmark{3},
     and Bruno Lepri\IEEEauthorrefmark{3}
     }
 \IEEEauthorblockA{
     \IEEEauthorrefmark{1}Department of Geography \\
     King's College London, London WC2B 4BG, United Kingdom\\ 
     Email: michele.ferretti@kcl.ac.uk
     }
 \IEEEauthorblockA{
     \IEEEauthorrefmark{2}Department of information engineering and computer science,\\ University of Trento, Trento, Italy
     }
 \IEEEauthorblockA{
     \IEEEauthorrefmark{3}Fondazione Bruno Kessler, Trento, Italy\\
     Email: barlacchi@fbk.eu\\
     Email: llucchini@fbk.eu
     }
 \IEEEauthorblockA{
     \IEEEauthorrefmark{4}ISTI-CNR, Pisa, Italy\\Email:lucapappalardo@isti.cnr.it}
 }
%  ---------------------------------------------------------

\maketitle

% TODO: RIGUARDARE ABSTRACT
\begin{abstract}
 The availability of massive data-sets describing human mobility offers the possibility to design simulation tools to monitor and improve the resilience of transport systems in response to traumatic events such as natural and man-made disasters (e.g., floods, terrorist attacks, etc\dots). In this perspective we propose {\scshape Achilles}, an application to models people's movements in a given transport mode through a multiplex network representation based on mobility data. {\scshape Achilles} is a web-based application which provides an easy-to-use interface to explore the mobility fluxes and the connectivity of every urban zone in a city, as well as to visualize changes in the transport system resulting from the addition or removal of transport modes, urban zones and single stops. Notably, our application allows the user to assess the overall resilience of the transport network by identifying its weakest node, i.e. \textit{Urban Achilles Heel}, with reference to the ancient Greek mythology. To demonstrate the impact of {\scshape Achilles} for humanitarian aid we consider its application to a real-world scenario by exploring human mobility in Singapore in response to flood prevention.
\end{abstract}


\section{Introduction}
\label{sec:problem_statement}
Cities are a physical manifestation of continuous processes of human interaction and of the weaving and knitting of social relations unfolding through space and stratifying through time. Cities are inherently ``hard'' to understand; their interlacing facets and tangled assemblages render them difficult to frame \cite{Batty2008, kitchin2011code, Kelley2014}. In particular, the nexus of human mobility patterns, transport planning choices and urban policies pose great difficulties for scientific modelling. It presents, in fact, what Rittel \emph{et al.} \cite{Rittel1973} call a ``wicked problem'', that is, a problem whose social nature and lack of objective truth makes inherently hard to tackle with traditional analytic tools. 

Despite being still far from adequately composing these problems \cite{Batty2008}, nowadays the  variability of massive data describing human movements allows planners and policy-makers to face relevant urban challenges through computational methods \cite{dedomenico2015personalized, rossi2018modeling,denadai2016death,barlacchi2017structural,pappalardo2015returners,barlacchi2015multi,ahmed2016multi,pappalardo2018data,Barlacchi2017}. While the observation of mobility flows offer for instance the possibility to investigate the inner workings of urban transport systems, this is not enough for efficient transport planning. As Ganin \emph{et al.} \cite{Ganin2017} note, traditional transport studies focus on efficiency measure by observing a transport network under normal operating conditions. This approach though fails to capture information that really matters for real-world applications, i.e. how the network reacts when is under stress conditions. Such is the case of traumatic events that might cause disruption of service or worse, like in the event of natural or man-made disasters (e.g., floods, terrorist attacks, etc.). Thus, uncovering weak points that affect a network's resilience, a proven key metric for transport policies \cite{Ganin2017}, is not only a task of paramount relevance for the design of proper intervention scenarios in case of disasters, but also an actual requirement for managing transport services in complex global cities like Singapore. We refer to such weak points as \emph{Urban Achilles Heels.}\footnote{https://en.wikipedia.org/wiki/Achilles\%27\_heel}

Starting from these considerations, we address the following questions: \emph{(i)} What and where are the weakest transport routes in a city, i.e., its \emph{Urban Achilles Heels}? \emph{(ii)} Given some changes in the transport system, what scenarios are likely to occur and what is their impact on human mobility and on the resilience of a transport system? 

Here we propose {\scshape Achilles}, a web-based application which provides an easy-to-use interface to explore the mobility fluxes and the connectivity of urban zones in a city, as well as to visualizes changes in the transport system resulting from the addition or removal of transport modes, urban zones and single stops. {\scshape Achilles} is based on a multiplex network representation of mobility data \cite{dedomenico2014navigability}, where every layer describes people's movements with a given transport mode, e.g. buses, metros, taxis. A node in a layer represents an urban zone of the city, edges indicate routes between zones and edge weights indicate the amount of people moving between two nodes in a given time window. {\scshape Achilles} exploits this network representation and allows the user to visualize changes in the transport system resulting from the addition or removal of transport modes, urban zones and single stops. Notably, {\scshape Achilles} allows to compute the \emph{Heel-ness} of a zone, a measure introduced in this paper which indicates the probability of a zone to disconnect the network when edges adjacent to it are removed. The computation of the \emph{Heel-ness} allows {\scshape Achilles} to state that the transport system has an \emph{Urban Achilles Heel} if there is at least one urban zone with a non-zero \emph{Heel-ness}.
%From the multiplex network we extract a set of measures indicating a layer's carrying capacity and its ability to satisfy the overall mobility needs in a given urban area \cite{barabasi2016network}. 

We show how {\scshape Achilles} can be used to explore human mobility in Singapore and test the resilience of its network to natural disaster, by using mobility data from several sources. The application allows to point out weak routes among urban areas in the city (i.e., routes where public transport does not meet the needs of the citizens), and simulate changes in the capacity of public transport to satisfy needs of citizens when specific events occur in the city, e.g. closing/adding transport modes or subway/bus stops. Additionally, it can be used by supplement humanitarian agencies for resilience preparedness and response, and in general fragile contexts where there is a dire need of deploying fast, scalable and effective humanitarian aid. Our approach is in fact highly flexible since it uses only data about transport and mobility flows. Given that many open datasets of such nature are nowadays publicly available\footnote{https://data.gov.sg/group/transport}, our approach can be potentially applied to any other urban area to simulate traffic changes due to specific events, such as the impact of adding or removing transport modes or stops, the impact of closing the access to an urban area, or the organization of city-wide public events.

\section{Related Work}

% EXPAND WITH SINGAPORE DESCRIPTION?
% la metto qui perche' e' la sezione con meno roba
% Despite Singapore's renowned efficiency, its transport services still face daily challenges which might undermine its economy and negatively impact the well-being of its inhabitants. 
%Common problems in the public transport system are related to a non-optimal positioning of bus stops or subway stations; to prolonged waiting times at such stops; or to misaligned interconnections and inter-modal routes between different transport networks (e.g., subway and bus). 
Understanding human mobility is a long-standing and highly relevant challenge for today's increasingly urbanized world \cite{UnitedNations2014}. In their reconstruction of the discipline's evolution Barbosa-Filho \emph{et al.} \cite{Barbosa-Filho2017} highlight how through time different scientific communities attempted to frame it using disparate data sources; starting from geographers' studies of spatial interaction and regional migration \cite{zipf1946p, hagerstrand1970people}. It is only recently though that significant advancements having been made due to novel data sources, such as Call Detail Records (CDRs) and GPS location data collected through mobile phone devices \cite{gonzalez2008understanding, pappalardo2015returners} and analyzed through the lens of network science \cite{simini2012universal,barthelemy2016structure}.

Using precisely methods from network science we can design powerful models and simulation tools for what-if analysis of different urban planning scenarios \cite{pappalardo2016human, jiang2016timegeo}. Singapore, due to the complexity of its transport system, is precisely a global city where such tools would be particularly valuable  \cite{jiang2017activity}.
Network science applied to the study of transport systems has proved to be increasingly useful in recent years.  Xu and Gonz\'alez \cite{xu2017collective} have shown that a slight re-routing of a fraction of daily car commutes produces significant reductions in traffic, thus alleviating the congestion state of the overall transport system. In this perspective the development of data-driven analytic tools can help control the transport system and improve both the customer's travel experience and the system's overall efficiency.  {\c C}olak \emph{et al.} \cite{colak2016understanding} have investigated the interplay of number of vehicles and road capacity to determine the level of congestion in urban areas. In particular, they show that the ratio of the road supply to the travel demand can explain the percentage of time lost in congestion. De Domenico \emph{et al.} \cite{dedomenico2014navigability} show that the efficiency in exploring the transport layers depends on the layers' topology and the interconnection strengths. 

Although these works doubtlessly shed light on interesting aspects about the structure of urban transport, they do not provide easy-to-use tools for exploring a city's demand for mobility and the efficiency of the related transport system. Giannotti \emph{et al.} \cite{giannotti2011unveiling} partly overcome this problem by proposing a querying and mining system (M-Atlas) for extracting mobility patterns from GPS tracks. This system, still, does not provide a method to assess weak points in cities, those that will collapse first in case of traumatic events.  

Wang \emph{et al.} \cite{Ip2009} highlight in fact many studies in resilience engineering that try capture different facets of the problem at stake, often tackling the most disparate domains (e.g., from spatial economics \cite{reggiani2002resilience} to computer networks \cite{scheffel2006optimal} and socio-ecological systems \cite{baggio13708socioecological}). However, as pointed out by \cite{Ganin2017}, since the majority of transport-oriented policies stress efficiency aspects measured under optimal conditions, and neglect to consider instead stress scenarios that happen in the real word, there is still a need to provide policy-makers with adequate tools to actually capture a transport network resilience; here intended once again as the network's capacity to recover from a catastrophic event returning to its original working state. Moreover, there is still a need to design tools to test and prototype scenarios for rapid response under critical conditions.

\section{Data Sources}\label{data-sources}
In this work, we use different data sources from the city of Singapore with information about (i) transport, (ii) administrative boundaries, (iii) mobility flows and (iv) floods.

\paragraph{Transport Data}
The public transport data are provided by LTA Data Mall\footnote{https://www.mytransport.sg/content/mytransport/home/dataMall.html}, which published a wide variety of land transport-related datasets both static and in real-time. In this work we use data about bus lines\footnote{https://www.mytransport.sg/content/mytransport/home/dataMall.html}, which provide information to build a transport network describing the displacements of inhabitants between different urban zones of Singapore. In particular, for every bus line, we retrieve information about its stops and the GPS traces describing the bus route. 

\paragraph{Administrative boundaries}
The urban zones areas are provided through the \emph{shape-files} at different administrative division levels of the city, and are available at the website \url{data.gov.sg}\footnote{https://data.gov.sg/dataset/master-plan-2014-subzone-boundary-web}. We use the most fine-grained division and, by applying a spatial join operation, assign every bus stop to the corresponding urban zone. The resulting dataset contains 323 urban zones and 4856 bus stops. 

\paragraph{Mobility flows} We use data indicating both the presence of people in every urban zone and the fluxes of people, i.e. Origin-Destination (O-D) matrices \cite{jain1999estimating} between urban zones at a given date and time.   
Such data were obtained from the data-as-a-service APIs provided by DataSparkAnalytics\footnote{https://www.datasparkanalytics.com/products/data-as-a-service}. In particular, the presence of people and the O-D matrix are estimated starting from the mobile network data provided by different Asian telco operators. It is worth mentioning that location signals from the telcos are analysed in an anonymous and aggregated way. We use these data to estimate the number of people moving by bus between two urban zones in a given time window, since official information about the number of users traveling on the buses is not available. We aggregated the flows by hour.

\paragraph{Flooding data} Flood prone areas forecasts, computed by the Singapore's Public Utilities Board (PUB), are provided as an incentive to reduce the risk of recurring flood events and to foster proactive measures among developers and the general public. We used such areas, retrievable at \url{www.pub.gov.sg}\footnote{https://www.pub.gov.sg/drainage/floodmanagement} to inform our use-case with substantiated data related to a real-world problem related to a specific setting.

We hence obtain for every urban zone $z$: (i) the bus stops in $z$, (ii) the bus lines passing through $z$, and (iii) all the urban zones connected to $z$. We define two urban zones $z_1$ and $z_2$ to be connected if there is at least one bus line connecting $z_1$ and $z_2$.

%In Figure \ref{coll-degree} we show the collapsed degree distribution of the urban zones in Singapore. In particular, it shows the distribution of the number of areas that are connected each other. [LORE: aggiusta questa definizione :)]
% [METTIAMO QUI COME RIDISTRIBUIAMO GLI UTENTI SULLE LINEE?]

% \begin{figure}
% 	\centering
% \includegraphics[width=0.5\textwidth]{imgs/degree_distro.pdf}
% \caption{Collapsed degree distribution of urban zones in Singapore. }
% 	\label{coll-degree}
% \end{figure}

It is worth to mention that despite our experiment being limited to the bus network system, due to limited data availability, further extensions to truly multimodal systems can be integrated with ease. This will enhance the modelling of transport planning scenarios with increased reliance and confidence. Limitations standing, our proposed tool can already serve for specific use-cases, such as precisely the considered instance of Singapore's bus network resilience to floods.

\section{Methodology}\label{methods} 
The use of complex networks to analyze complex systems is a powerful way to simplify the description of a problem by means of a process of abstraction. This process reduces the complexity of the structure to two simple elements: nodes, i.e. the objects of the system, and edges, i.e. the relations between these objects. However, such a representation can lead to a strong loss of information about the real system characteristics. 

In this context, an interesting novel approach was proposed by De Domenico \emph{et al.} in their seminal work on multilayer networks \cite{dedomenico2013multilayer}. They have proposed a new framework in which it is possible to encode information about the different possible relations between the nodes of the system in different layers. Each of these layers represents a specific kind of relation between a set of nodes and can be considered as a partial representation of the whole system without the loss of a simple structure representation: for example, a sub-network describing a specific kind of relation between the nodes of the network, which again reflects some of the properties or the characteristics of the complex system. In this work we adopt a similar mathematical formulation to describe the transport system of an urban environment. To better describe the transport system of the city we are studying, in Section \ref{casestudy} we specifically model the network structure to best fit the city characteristics following the more general formulation discussed herein.

In an urban area, two zones can be connected through several transport means (e.g., bus, taxi, subway, etc.). As each transport mean differs from the other, also each subway or bus line differs from the others. In fact, using two different lines of the same mean require a transfer at some stop. In turn, each of these transport modes may be considered as a set of different lines following a specific path; connecting specific zones of an area; and eventually, can be considered as an additional layer of the system. To express this kind of information we introduce the concept of urban multiplex network by means of the following definition:\\

\paragraph*{Definition} An \textbf{Urban Multiplex Network (UMN)} is a network in which two nodes represent zones of an urban area and can be connected, at the same time, by multiple edges that belong to different \emph{layers}. We model such structure with an edge-labeled multi-graph denoted by $G = (V, E, L)$ where $V$ is a set of nodes, i.e. the zones in which the area we are studying is divided; $L$ is a the set of layers which encode the information about the different possible types of connections between two nodes, e.g. the public transport mean characterization and/or the line number; $E$ is a set of labeled edges, i.e., a set of triples $(u, v, l)$ where $u, v \in V$ are two nodes, and $l \in L$ is a layer of the multiplex network.\\

The global connectivity of two zones, which is the multidimensional connectivity (i.e., the connectivity considering all the layers of the network) in an urban area is a combination of two elements: (i) \emph{connection intensity} and (ii) \emph{connection redundancy}. We define the intensity of the connection between two zones on a single layer as:\\

\paragraph*{Definition} \textbf{Connection intensity}\\
\begin{equation}
h_l(u, v) = w_l(u, v) \frac{|\Gamma_l(u) \cap \Gamma_l(v)|}{min(|\Gamma_l(u)|, |\Gamma_l(v)|)},
\end{equation}
where $w_l: V \times V \times L \rightarrow \mathbb{N}$ is a weight function representing the mobility flux between two zones on layer $l$, and $\Gamma_l(u)$ is the set of neighbours of the zone $u$.
Connection intensity consists hence of two factors: the first factor, $w_l$,  indicates the percentage of people moving between the two zones using a specific transport layer and line $l$. The second factor, $\frac{|\Gamma_l(u) \cap \Gamma_l(v)|}{min(|\Gamma_l(u)|, |\Gamma_l(v)|)}$, is the percentage of common neighbours, ${|\Gamma_l(u) \cap |\Gamma_l(v)|}$, with respect to the most selective zone, $min(|\Gamma_l(u)|, |\Gamma_l(v)|)$. The idea is that, on each layer, the connection intensity is influenced by the number of displacements between the two zones, weighted by the value of \textit{selectiveness of the most selective zone}, i.e. the ratio between the neighbours shared by the two zones and the number of neighbours of the zone with the smallest set of neighbours.\\

The second element on which global connectivity depends is connection redundancy, which takes into account the relevance of a layer in the all-layers zone's connectivity. This quantity measures to what extent the removal of the links belonging to a specific layer affects the capacity to reach a neighbouring zone from the considered one. Redundancy in general is a measure of the strength of node-node connections considering the links connecting the two nodes over all the layers of the network. Before defining the connection redundancy, however, we need to introduce the layer $l$ relevance, $LR(u,l)$, for the node $u$.
$LR(u,l)$ can be defined as the ratio between the number of nodes that disconnect from node $u$ if all the links of layer $l$ are removed (which is the same as removing an entire layer to the multiplex network) and the number of nodes reachable before the removal of the layer\cite{berlingerio2011foundations}:\\

\paragraph*{Definition} \textbf{Layer Relevance}\\
\begin{equation}
LR(u, l)=1-\frac{N_r(u,l^c)}{N_r(u,L)}
\end{equation}
where $N_r(u,L)$ is the number of reachable nodes from node $u$ using all the multiplex network's layers $L$, while $N_r(u,l^c)$ is the number of reachable nodes from $u$ after the removal of layer $l$, i.e. the reachable nodes from $u$ considering only the set of layers $l^c=L\setminus l$.\\

\paragraph*{Definition} \textbf{Connection Redundancy}\\
\begin{equation}
r_l(u, v) =(1-LR(u, l))\cdot(1-LR(v, l))\cdot\sum_{m=1}^{L}\frac{A_{u,v,m}}{L},
\end{equation}
where $A_{u,v}$ is a vector element of the adjacency tensor of the multiplex that keeps track of the presence of edges between node $u$ and node $v$ over the $l\in L$ different layers.
We give a higher score to the edges that appear in several layers, so we are interested in the complement of those values. If the two areas are linked in more than one layer, the score raises until a maximum of 1.\\

We combine connection intensity and connection redundancy taking into account the multidimensionality of multilayer network's connectivity: a greater number of connections on different layers is reflected in a greater chance of having a strong connectivity.\\

\paragraph*{Definition} \textbf{(Global) Connectivity.}\\
Let $u, v \in V$ be two nodes and $L$ be the set of layers of an urban multiplex network $G = (V, E, L)$. The connectivity of two urban areas $u, v$ is defined as:
\begin{equation}
c(u, v) = \sum_{l \in L} h_l(u, v)(1+r_l(u, v)).
\end{equation}

It is worth noticing that the measure proposed can be used to estimate the strength of the connectivity between nodes also in single-layer networks, in which $r_l$ is one and the overall sum is $h_l$.\\

The aim of this work is to provide policy-makers, transport planners and humanitarian agencies with a working tool that has the versatility to adapt to different urban areas and the capacity to quantify the level of susceptibility to disruptive events related to a transport system own structure. More specifically, we are interested in studying the weaker ties as those potentially responsible for the disruption of the multiplex network. In particular, to study the weaknesses of a specific network, we introduce the notion of \textit{Heel-ness} as the relative importance of a node in the context of network navigability.\\

\paragraph*{Definition} \textbf{Heel-ness.}\\
Given a zone $u \in V$, the \emph{Heel-ness} of $u$ is defined as:
\begin{equation}
H(u) = max\bigg[\frac{|V-R_{u-(u,v,l)}|}{min[c(u,v)]_{v\in N_u}}\bigg]_{l \in L},
\end{equation}
where $N_u$ is the set of neighbors of $u$, $R_u$ is the set of reachable nodes from node $u$ while $R_{u-(u,v,l)}$ is the set of reachable nodes from node $u$ after the removal of the edge $e = (u,v,l) \in E$. As a consequence of these definitions  $|V-R_{u-(u,v,l)}|$ is the number of non-reachable nodes after the removal of the edge $e$.\\

This metric quantifies this importance in terms of optimal disruptive events classified following the Granovetter's theory for single-layer networks \cite{granovetter1973weakties}: the importance of a node in preserving a network's giant component depends on the connectivity of the node. In particular, nodes which shares weak connections tend to be those more important to be preserved. Heel-ness measures the importance of a node considering the weakest connection it has, and the number of unreachable nodes after that edge disruption.
The Granovetter's theory was proposed in 1973 in the context of social networks. Since then it was studied and proposed in countless different fields of application. Following a similar approach we adapt the measures introduced in \cite{pappalardo2012how} to detect important ties in transport networks. In particular we focus our attention on the topological resilience of multilayer networks as proposed in \cite{dedomenico2014navigability}. Topological resilience studies the robustness of a network in resisting to topological failures, such as edge removal. Heel-ness measures the single-event effects of the network and ranks them depending on the greatest damage they can inflict to the network structure if one particular edge is removed. Extending then the idea from \cite{granovetter1973weakties} to transport networks allows us to refer to weak ties as those edges that make the network more fragile and susceptible to targeted edge loss. We stress further that topological resilience takes into account only the network structure while edges can be weighted in such a way that more important links, e.g. in a transport network those links that sustain higher traffic flows, are naturally considered as more crucial for the overall system connectivity.

Being able to classify nodes depending on the importance of their connection in the network, we can classify networks based on the presence or absence of fragile nodes. Using the definition of Heel-ness we can state that an urban multiplex network has an \emph{Urban Achilles Heel} if there is at least one node for which the \emph{Heel-ness} has a non-zero value.\\

\paragraph*{Definition} \textbf{Urban Achilles Heel.}\\
We define the \emph{Urban Achilles Heel} of a multiplex network as the node $h \in V$ for which 
\begin{equation}
	H(h)>0 \hspace{.1cm}\text{and}\hspace{.1cm} H(h)\geq H(u) \hspace{.25cm}\forall u \in V.
\end{equation}

This definition provides a classification based on a first-order disruptive event, i.e. an event that only causes the loss of a single edge. However, the extension of this definition to further orders is straightforward and can be done by sub-sequential removal of the $n$ weakest ties of the network.\\

Applying this definition to real transport networks can help in the identification of important edges and nodes that are crucial for the stability and routing of all zones in an urban area. 

\section{Application Design}
%The overall application is freely available online at \url{http://achille.fbk.eu}. For scientific dissemination and to provide access for a larger external non-technical audience we also provided a video-tutorial can be found at \url{http://achille.fbk.eu/about}. 
The application is freely available online at http://achille.fbk.eu. For scientific dissemination and to provide access for a larger external non-technical audience it was also complemented by a video-tutorial http://achille.fbk.eu/about. The application design follows closely the analytical framework described in Section \ref{sec:problem_statement}, while implementing a client-server architecture pattern. We implemented a RESTful web service using \emph{Flask}, a widely used micro web framework written in Python. In the REST architecture the server exposes the resources to a client trough a set of defined URLs. For example, in order to get the list of buses passing trough an area, the endpoint is defined as \url{http://achille.fbk.eu/area/<id_area>/services}. This API is responsible not only for data provisioning, but also acts as the communication layer between the user and the network models. It is worth noting that, given its modularity, our application can be re-purposed as an agnostic provider of services to other consumers. In particular, the exposed methods consist in:
%The back-end component of such structure is responsible for implementing  and serving the models presented in Section \ref{methods}.  It is a RESTful webservice
%The client application then consumes the services via a RESTful web services service exposed by the same server. The webservice is based on \emph{Flask}, a widely used micro web framework written in Python. 
\begin{itemize}
%\item a forecasting endpoint  for predicting the movement of people in a Singaporean urban zone, based on the historical data provided by the local transport authority;
\item a query endpoint returning  the network features' geometries for a given urban zone ID, and additional information used to populate the geographic map application;
\item a second query endpoint returning for each given urban zone ID the computed network metrics, i.e., \emph{Connection intensity, Connection Redundancy, Connectivity} (Section \ref{methods}).
\end{itemize}

\begin{figure}
	\centering
\includegraphics[width=0.5\textwidth]{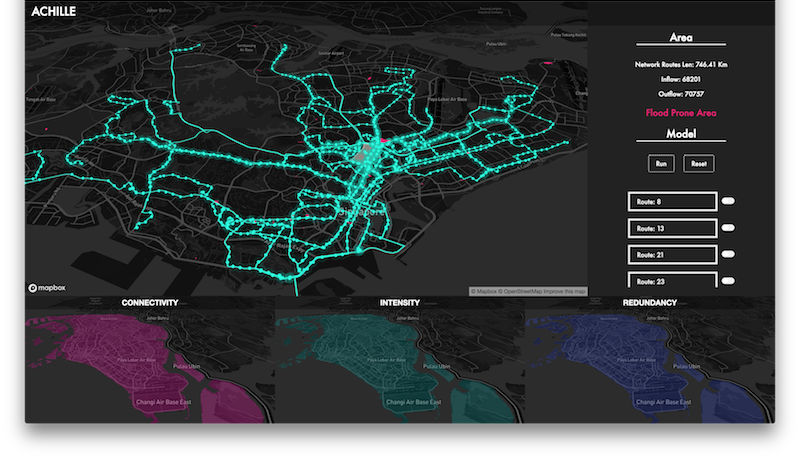}
\caption{Application User Interface (UI): Upon an area selection, the user is presented with the 3D network metrics for the city of Singapore. The interactive menu on the right allows then to select and deselect transport routes.}
	\label{app-img}
\end{figure}

\begin{figure}
	\centering
\includegraphics[width=0.45\textwidth]{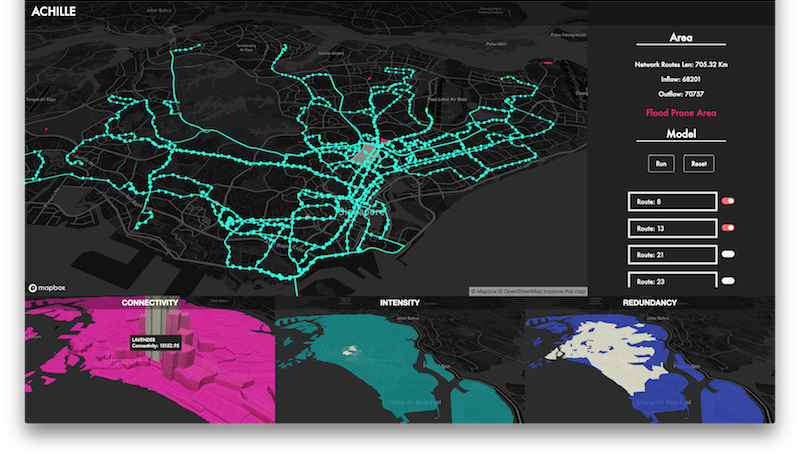}
\caption{Metrics: The ``Run'' button allows to calculate the intensity, redundancy and connectivity measures and to visualize them in the three bottom windows. Further, every interaction with the bus lines menu will trigger the computation of the Heel-ness metric. In the above image, four areas are identified as such vulnerable sites.  }
\label{city_panel}
\end{figure}

\begin{figure}
	\centering
\includegraphics[width=0.45\textwidth]{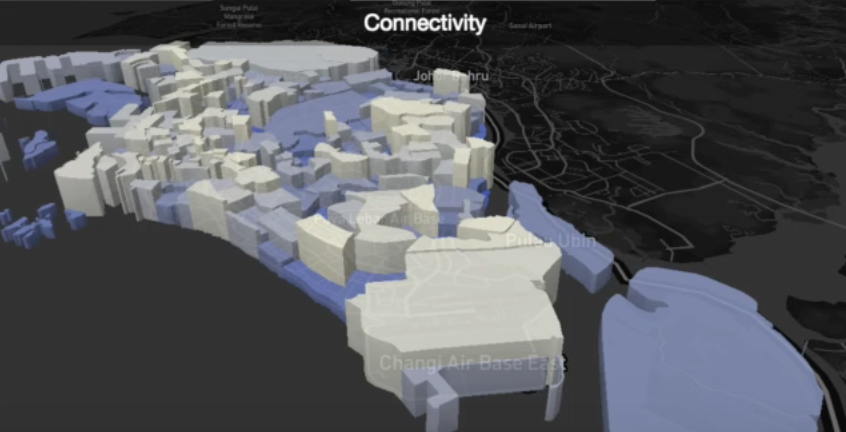}
\caption{Zonal Focus (example): The 3D maps in the rightmost bottom window inform the user with additional metrics. In the example, the height of the extrude zone is proportional to the average value of its multidimensional connectivity, computed across all the connected urban zones.}
	\label{window}
\end{figure}

The front-end application is thus the entry-point for quickly interacting and prototyping future transport scenarios. The User Interface (UI), visible in Figure \ref{app-img}, has been developed with easiness of use and clarity of interpretation its standard pillars. It allows a non-technical audience to inspect the number of routes connecting an urban zone to the rest of the city simply by clicking on an urban zone in the city panel and control them via an interactive menu (Figure \ref{app-img}). Upon addition/removal of one or more routes in the menu, the user can trigger the calculation of the network metrics (Figure \ref{city_panel}), which are promptly displayed in three separate windows. Every window shows a 3D map of the city, where an urban zone's height is proportional to its average networks value computed over the connected urban zones (Figure \ref{window}). The menu and the bottom windows allow the user to simulate how the city's connectivity changes after, for example, the construction of a new route or the temporary closing of an existing one. Further, the identification of sensitive areas, the so-called \emph{Urban Achilles Heel}, can be accessed through a revealing button. The example in Figure \ref{city_panel} presents four areas in the northern part of Singapore.

All of the geographic map components are fully interactive and built with the latest web-mapping technologies taking advantage of latest WebGL standards and 3D capabilities; as such, they allow for a most fluid and seamless experience. This is a crucial factor that allows the application to hide the complexity of the models behind a fluid interface. It also let the technological stack move out of the way, making room for discussions and human elements that enhance decision-making processes.

\section{Case study: Floods prevention in Singapore}\label{casestudy}
In the following section we discuss the specific modeling adopted for the Singapore's transport network and the results obtained by studying the structure of the system and its resilience to synthetic disruptive events.

\subsection{Network structure}
When dealing with real-world problems a case-dependent modelling usually ensures better performances and results in terms of actionable solutions. In the advanced case study, we deal with Singapore's transport network with the aim to specifically uncover possible solutions to disruptive events (e.g. intense rainfall and consequent floods\dots) which might negatively impact the road network. We apply the methodology presented in section \ref{methods} considering the reduced transport network composed by more than 300 different scheduled bus lines. This operation enable us to scope down the issues and focus solely on those related to the road network. 

\begin{figure}
	\centering
\includegraphics[width=0.45\textwidth]{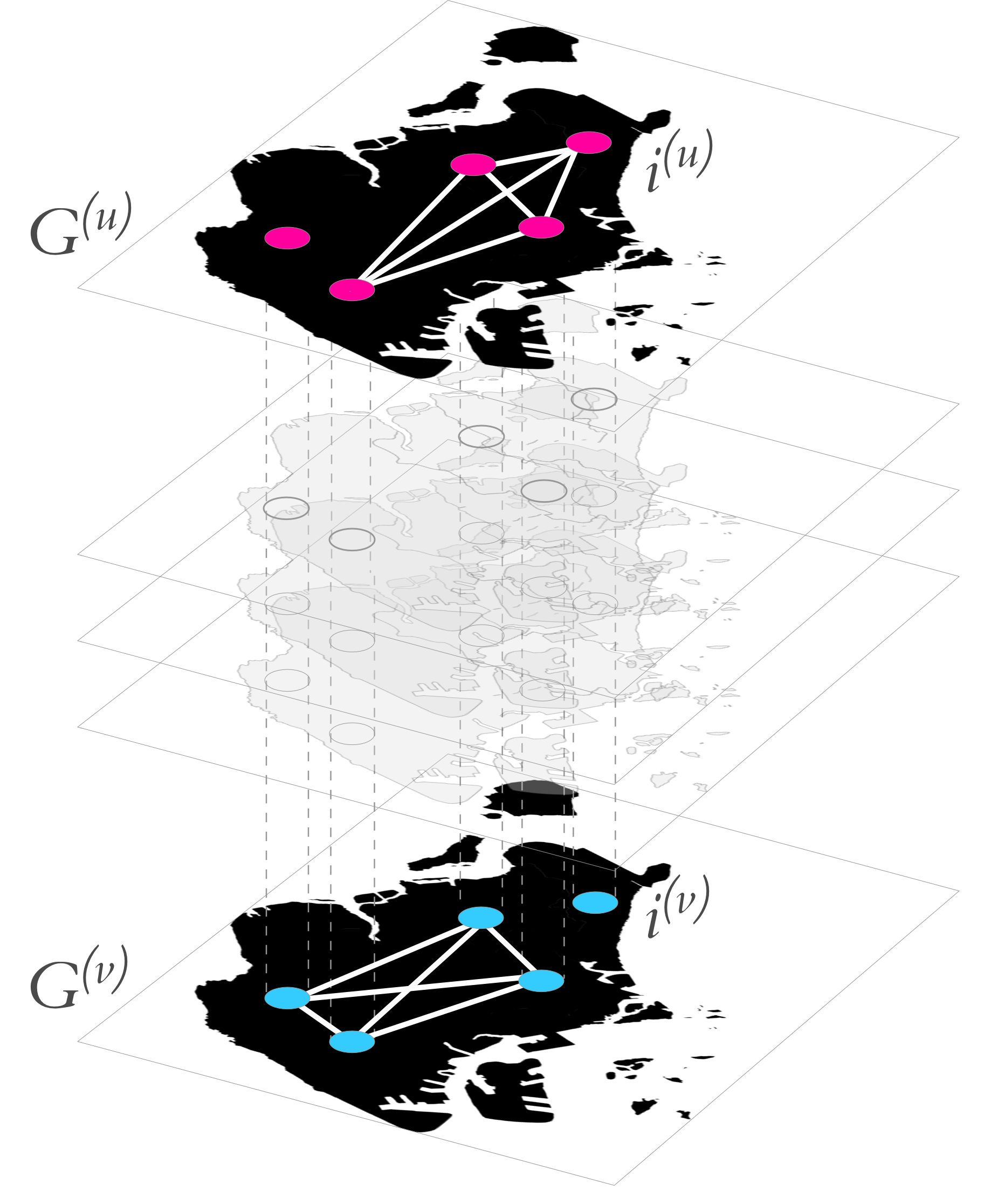}
\caption{Schematic of the structure of the multiplex network adopted to model the road public transport services for Singapore. Each layer represents a different bus line. Each of these fully connects different zones of the city. This structure naturally takes into account possible rerouting strategies that can maintain the layer partially connected.}
	\label{multiplex}
\end{figure}

In our framework, each bus line represents a layer of the network; each layer consists of all the Singapore's administrative areas (according to a government sanctioned classification available at \url{data.gov.sg}\footnote{https://data.gov.sg/dataset/master-plan-2014-subzone-boundary-web}); each zone is a node of the multiplex. Further, each bus line fully connects to every zone in which a bus line has a stop with all the others zones in which the same bus line has a stop. In this configuration, disruptive events that involves only limited zones, which may not necessarily disconnect the entire line if a reroute is prompted, are taken into account. The  model is indeed built in such a way that edges can be removed gradually depending on the nature of the disruptive event we want to simulate.
In figure \ref{multiplex} we schematically represent the network structure used to model and analyze the Singapore's bus transport network. As depicted, each layer share with the others the same nodes but changes the connections between the zones depending on the bus line it represents.

Even in this specific network configuration, the metrics introduced in section \ref{methods} maintain their significance. The relevance of each layer depends, in fact, on the relative number of neighbours in a layer with respect to the multiple degree of the node. In particular, the shape of the connectivity, implemented from the definition proposed in \cite{berlingerio2011foundations} ensures that the Heel-ness of a node consistently depend on the node's layer relevance and the node's connection redundancy instead that only on the connection intensity. We stress that, if this would have been the case, fully connected layers would have brought to an initial connectivity measure that would have encoded only operational information, i.e. information about the mobility flow passing trough links, for all the nodes that share the same bus lines independently on the possible different size of the disconnected component they might cause after a disruptive event involving one or all of their connections. This highlight the importance of a combined approach that considers both the structure of the multilayer network, i.e. relevance and redundancy of the connections, and the network usage, i.e. the information about mobility flows over different links.

\subsection{Experiments}
\begin{figure}
	\centering
\includegraphics[width=0.45\textwidth]{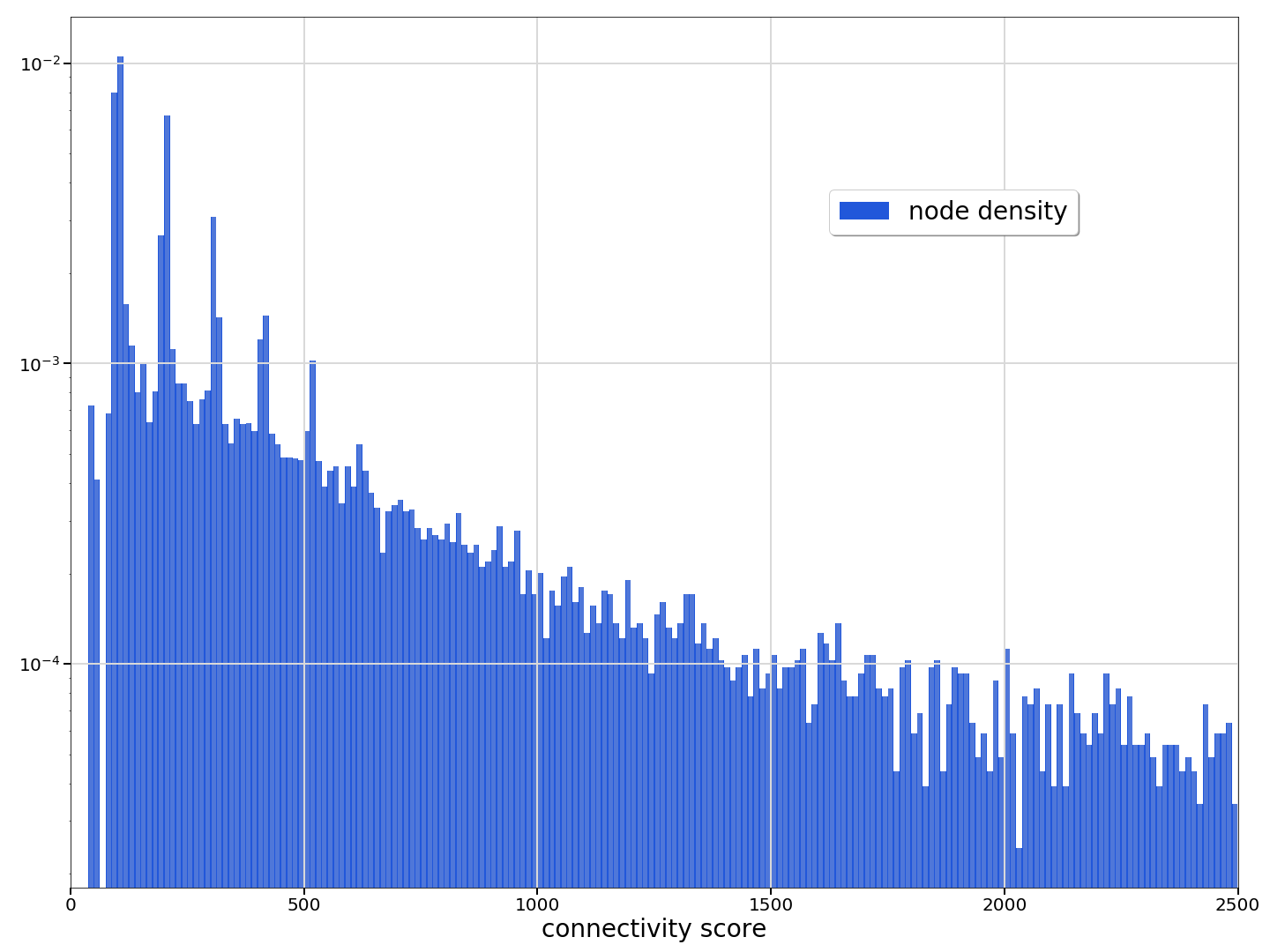}
	\caption{Connectivity value density within the network's initial configuration state. Connectivity is computed for each couple of nodes, which are then are sorted depending on the computed score.}
	\label{fig:connectivity_density}
\end{figure}

Moving from of the aforementioned considerations we conducted an extensive connectivity assessment by observing Singapore's urban network resilience to the removal of high and low connectivity links separately. 
We use this case study to test the behavior of multilayer node connectivity, as defined in \ref{methods}, on a real transport network. To this extent we compute the connectivity for every node in the network and we sort them accordingly to the obtained score. In figure \ref{fig:connectivity_density} we show the connectivity for the network in its initial configuration. We can see that there are two dominant behaviors. 

The left side of the histogram consists of areas with a low connectivity score. These are the areas dominated by the contributions given by the number of layers who share a connection showing different peaks. Each peak corresponds to the different number of shared layers. The right side consists of those urban zones that are geographically central and functionally important urban nodes; the number of layers that two such nodes share is no more dominating the contribution to the connectivity score, while its increasing importance is given by the length of the different bus lines and number of nodes they are connected with. 

As illustrated in Figure \ref{fig:connectivity_density}, connectivity can be used to classify and sort nodes depending on the strength and importance of their ties. This allows us to test the weak ties theory for transport networks. 

We start by removing the weakest tie in the network, and the subsequent in decreasing order of weakness, iterating the computation after each removal. The expected outcome is a faster network percolation than achieved adopting an opposite strategy, i.e. removing the stronger ties of the network first. The result of this experiment are presented in Figure \ref{fig:plot}. The relative size of the largest network component changes with the removal of a given percentage of links. The values are sorted by increasing (red solid line) or decreasing (blue dashed line) connectivity-score order. The red line shows the different speed of percolation faced by the same system under targeted attacks with different strategies. 

Additionally, it is worth noting how the network faces a first significant disruption after the removal of more than 50\% of total number of edges. This finding holds even with the adoption of a more effective disruptive strategy. Such result highlights the high degree of resilience exhibited by Singapore's bus network, which is consistent with the city's highly efficient transport network \cite{barter2013singapore}. Further studies might benefit thus from replicating the above experiments in other locales with different transport systems, to check whether a similar pattern is present and how it behaves to disruptive events. 

\section{Discussion and Implications}
Our simulated experiment has been conducted, due to data availability, only on the bus transport layer. However, it is straightforward to extend our framework to multiple transportation systems (e.g., metro and train). Testing the network resilience to the aforementioned stress conditions in a truly multimodal perspective represents thus an important tool at disposal of transport planners to assess (i) the current state of the transport network; (ii) to plan future operations; (iii) and to keep running the system at its overall optimal capacity.

\begin{figure}
	\centering
\includegraphics[width=0.45\textwidth]{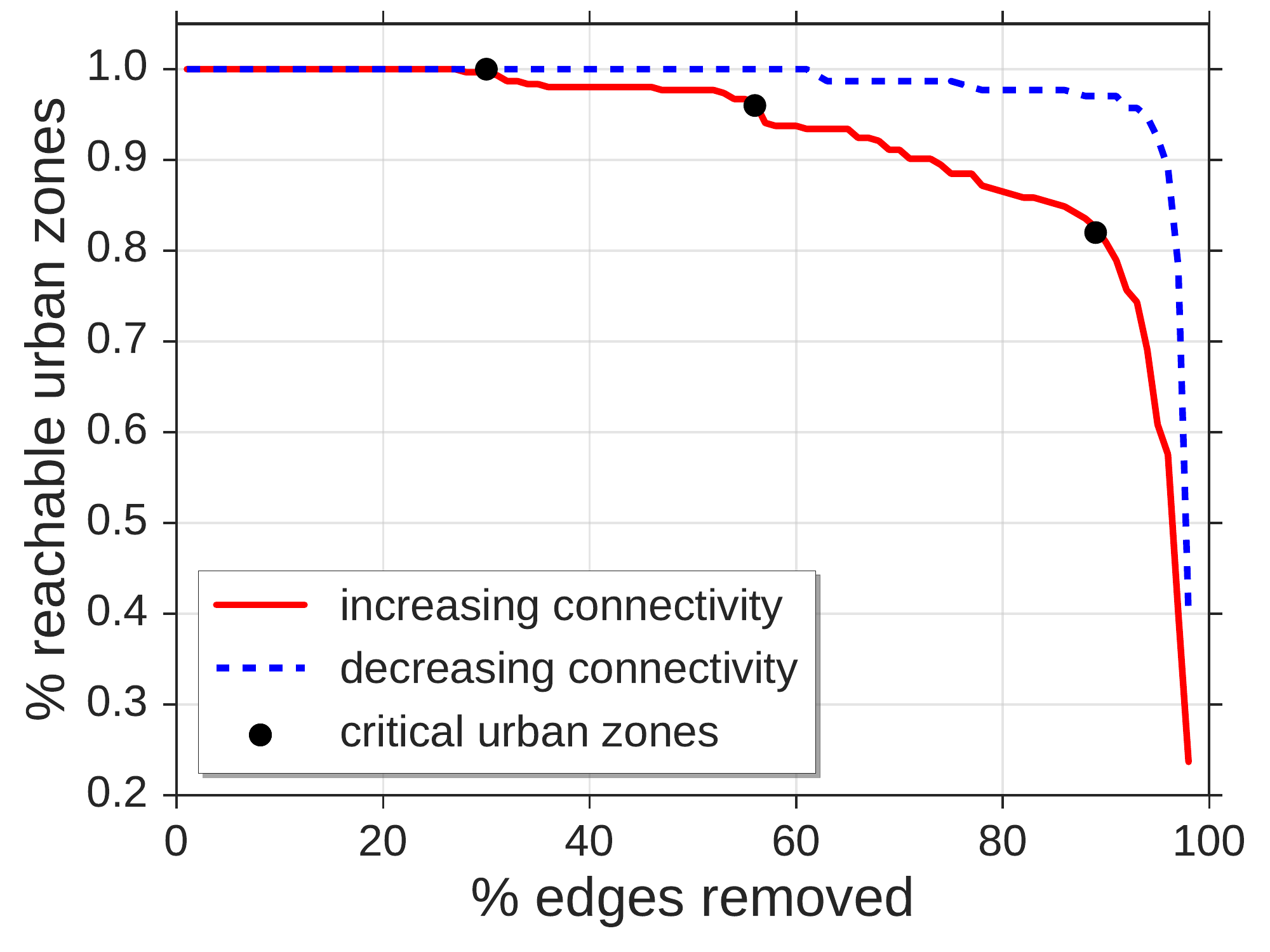}
	\caption{Stability of the urban network to links removal. The x axis shows the percentage of removed links. The y axis shows the size of the greatest network component.}
	\label{fig:plot}
\end{figure}

In particular, we found that Singapore has a resilient urban network, as all the nodes are still reachable when removing up to 30\% of the links. Moreover, the deletion of links in decreasing order of connectivity affect less the network's global connectivity, as more than 97\% of the urban zones are still reachable after the removal of 90\% of the links (Figure \ref{fig:plot}, blue dashed line). In contrast, when deleting the links in increasing order of connectivity the network crumbles faster, as almost 20\% of urban zones become unreachable after the removal of 90\% of the links (Figure \ref{fig:plot}, red solid line). These results suggest that the proposed connectivity and risk metrics can profitably be deployed to discover those urban connections whose existence is crucial for the network resilience. Moreover, those metrics can also help to evaluate the impact of changes and intervention in the city's mobility (e.g., the closure of a bus line). Figure \ref{fig:plot} highlights the points where the accelerated network disassembling commences, i.e., the \emph{Urban Achilles Heel}: up to those values the transport system exhibits a fair resilience, but surpassing such thresholds provokes a rapid network fall-out.

% ----------------------------------
% NUOVA SEZIONE PER I REVIEWERS:
% ----------------------------------
% "I think that a good part of the value of this contribution is the deployment of the application. However, from the description it is not clear to me how much the same interface could be replicated for other cities by independent researchers. In the ideal scenario, the authors should public a framework or service that, given in input a given type of data, it able to build the interface for any city. I am not sure the API the authors refer to does that--it seems that the API is specific to the Singapore data. I would suggest to provide more detail about this aspect."
% and 
% "- For example, it is not clear how scalable the application is to other cities. Did the authors develop this in such a way that maps and urban data from -say- London could be inserted?
Independently on the specific results obtained for the Singapore's transport network, we believe that our approach and our application can be used in several  situations, both from a geographical and structural point of view. 

\subsubsection*{Application deployment}
Considering for instance the level of complexity and the structural organization of Singapore \cite{barter2013singapore}, one of the most developed cities in the eastern hemisphere, our model simulation showcases a relatively sound resilience to potential catastrophic events. While this is an interesting result in itself, another valuable aspect of our application is its ease of reproducibility in urban contexts characterized by more fragile infrastructural systems. In cities and urban areas with newer or less developed transport systems the application of the proposed methodology might in fact yield higher impact results.   

From from a data availability standpoint our application can easily be deployed to other contexts. Out of the four major sources of information presented in section \ref{data-sources}, namely \emph{a) Transport Data, b) Administrative Boundaries, c) Mobility Flows and d) Flooding Data} only the very last one is specifically tied to the Singapore's context and the use-case introduced. Other data sources, such as \emph{Administrative Boundaries} can in fact be commonly found either through local authorities official data portals, or via open data alternatives such as \url{OpenStreetMap}\footnote{https://www.openstreetmap.org/}. The pivotal source, i.e. \emph{Transport Data}, can additionally be found across different urban settings also through open data standards such as the \emph{General Transit Feed Specification (GTFS)}, which  \emph{"\dots allows public transportation agencies to provide real-time updates about their fleet to application developers"}. GTFS is \emph{de facto} the standard for exposing and consuming public transit data; can be globally found and integrated through \emph{ad hoc} search engines such as \url{TransitLand}\footnote{https://transit.land/}; and come with a robust track-record of academic research and industry applications. \cite{ma2017delivering, ferris2009location, liu2017efficient}.

Lastly, in a final effort to extend the scalability of our application to different contexts and use-cases, we devise possible implications of our tool by distinguishing different typologies of users, namely urban planners and policy makers, and humanitarian agencies.

\subsubsection*{Urban planners and policy-makers}
Current approaches in transport network resilience focus on developing comprehensive frameworks or policies that largely prioritize efficiency and reducing delays, rather than considering the impact of stressful events on the network \cite{Ganin2017}. While these approaches are perfectly understandable from a policy's stand-point, they might often leave real-world transport systems vulnerable to changing conditions. These conditions might also be exacerbated in complex urban environments such as Singapore. Thus, the insights provided by testing such scenario might prove extremely useful instead in preventing catastrophic events and improving the overall resilience of the system.
% Operating in complex urban environments such as Singapore only  the issue. 

% For planners 
% The ability to quickly respond to failures and bring the system back to its carrying capacity in a timely manner is the crux of transport policies \cite{Ganin2017}

\subsubsection*{Humanitarian Organizations}
The possibility to prototype scenarios and test the mutable conditions of a system has direct implications for other cities; other kinds of disruptive events; and other sectors, such as the humanitarian response. The latter is in fact an area of application where the ability to prevent damaging phenomena by containing negative impacts and providing an adequate rapid response is absolutely crucial \cite{8167726}. By providing a clean interface and a tool able to scale over a wide portfolio of settings, our application is suited to address challenges where timing is the key factor, such as \emph{pre-event} preparedness prevention or \emph{post-event} rescue team deployment.

\section{Conclusion}

% Questo quello che abbiam fatto:
% il modello é semplice e versatile, implementa una misura di facile comprensione e applicazione; questo (insieme alla bella applicazione?) rende facile da utilizzare questo lavoro su diverse cittá anche da persone non esperte nel settore, in particolare per il caso di Singapore fornisce un risultato chiaro sulle maggiori fragilitá del sistema...queste fragilitá ci sono? nel video di cikm parlavate di una linea che fa nord sud...@gianni e @michele ...era vero? abbiamo dei numeri? 
Transport planning is a difficult arena to operate in. This is due to the inherently "wicked nature" of the issues at stake coupled with the complexity of urban environments \cite{Batty2008,Rittel1973}. Advancements in understanding human mobility have recently been made through the use of computational methodologies (see for instance \cite{colak2016understanding,pappalardo2015returners,barlacchi2015multi,ahmed2016multi}). However, the need to operationalize and distill such methods into actionable tools tailored to real-world problems is still present. In particular, the humanitarian sector might benefit from the application of models and methods coming from the new science of data. Our proposed {\scshape Achilles} application is a step forward in this direction, providing an easy-to-use interface on top of complex models that allow non-expert users, such as non-profit organizations, to rapidly prototype different stress scenarios. It represents thus a tentative example of how data science can be put to good use to tackle issues of social relevance. In the presented experiments we tested the case of Singapore, particularly suited due to its natural geography and the frequent disruption thereby caused on the bus transport network system by floods. Since the application is built deploying data that are not site-specific (e.g. O-D matrices), it is scalable to different environments, making it suitable to be applied even in fragile contexts. Further, due to its flexible architecture {\scshape Achilles} accommodates new data sources with ease, adapting its modelling capacity to meet local needs. As unpredictable stress events might present themselves in the future, transport planners need to be vary of mutable conditions. This requires not only a close monitoring of transport networks under normal operational settings, but the ability to quickly respond to new challenges \cite{Ip2009}. The solution we advance lies precisely in this direction.

\bibliography{biblio}
\bibliographystyle{plain}

\end{document}